
\magnification= 1200
\baselineskip=20pt
\tolerance=10000
\hfill {UR-1315    }

\hfill{ER-40685-765}

\hfill {IC/93/178}
\vskip 1cm


\def\dpfour{{d^4 p\over (2\pi)^4}}
\def\dq{{d^4 q\over (2\pi)^4}}
\def\oSigma{\bar{\Sigma}}
\def\oD{\bar{D}}
\def\od{\bar{d}}

\def\dkthree{{d^3 k\over (2\pi)^3}}
\def\pk{\vec{p}.\vec{k}}
\def\kr{ \vec {k} . \vec{ r}}
\def\pmais{(p_0+\omega_k)^2-\omega_{p+k}^2}
\def\pmenos{(p_0-\omega_k)^2-\omega_{p+k}^2}

\centerline{\bf OUT OF EQUILIBRIUM PHASE TRANSITIONS}
\centerline{\bf AND A TOY MODEL FOR}
\centerline{\bf DISORIENTED CHIRAL CONDENSATES}

\vskip 1cm
\centerline{Paulo F. Bedaque}
\centerline{and}
\centerline{Ashok Das}
\bigskip
\centerline{\it International Centre for Theoretical Physics}
\centerline{\it Trieste, Italy}
\centerline{\it and}
\centerline{\it Department of Physics and Astronomy *}
\centerline{\it University of Rochester}
\centerline{\it Rochester NY 14627}
\bigskip
\centerline{Abstract}

\bigskip
We study the dynamics of a second order phase transition in a situation that
mimics a sudden quench to a temperature below the critical temperature in a
model with
dynamical symmetry breaking. In particular we show that the domains of
correlated values of the condensate grow as $\sqrt{t}$ and that this result
seems to be largely model independent.
\vfill
*Permanent Address\hfill
\eject
{\bf I. Introduction}

Most of the studies in finite temperature field theory deal with problems in
thermal equilibrium. There are systems though, where conditions change so fast
that to take the equilibrium formalism, even as a first approximation,
 would be inappropriate.
One of the situations where
non equilibrium phenomena are interesting is the dynamics of phase transitions.
They have been considered by a number of authors, mostly in
connection with problems in cosmology. More
recently the interest in phase transitions ocurring
 in relativistic heavy ion collisions has grown due to the possibility of
producing in the laboratory
the deconfined phase of strong interactions. Clear signals that the
quark gluon plasma was indeed produced and the study of its
its properties is made
dificult by
the fact that in the laboratory this phase
is produced in a small region and is very
short lived.  Non equilibrium phenomena
are quite normal ocurrence in this kind of
experiment and this provides another motivation for considering the dynamical
aspects of phase transitions in relativistic field theories.

One very interesting proposal on how to decide whether the quark gluon plasma
was created or not during a heavy ion collision was suggested many years
ago and since then has been studied
by many authors [1]. It is based on the
fact that besides the deconfining phase transition,
we expect in QCD another phase transition at about the same temperature,
namely, the chiral symetry restoration. The picture goes as follows. The
value of
the order parameter $\bar\psi   \psi$ becomes zero at high temperatures
reached in the plasma formed by the colliding nuclei and chiral
symmetry is restored.  With the expansion of the plasma though, this
temperature
decreases and eventually chiral symmetry is broken again. The direction
of symmetry breaking  in
isospin space, however, does not necessarily have to be
the same as it originally was before the collision took place. Therefore,
 if the value of the order parameter is, say, in the
$\pi^0$ direction, after hadronization this plasma will be converted into
neutral pions only, and not charged ones. This
could lead to spectacular events
with large number of pions correlated in isospin space  coming out of one
collision and signaling clearly that chiral symmetry restoration was achieved
inside the plasma. Events of this kind  might have already
been seen in cosmic
rays (the so called Centauro events [2]) as it has been sugested [3]. The
dificulty
with this picture is that the direction in which chiral symmetry is broken does
not to have to be the same all over the plasma. More likely there will be
domains in which the order parameter is the same, but different domains
will have diferent symmetry breaking directions. This domain structure
is a very general phenomenon in condensed matter, cosmology, etc.
If the typical domains are small and only a few pions
 can be made with energy contained
in each of them, then the correlation in isospin space can hardly be
distinguished from that
of a random production. In equilibrium there are indications that this is
indeed the case. Recently Rajagopal and Wilczek [3] have suggested that if
the rapid cooling of the plasma is modelled as a sudden quench below the
critical temperature the size of correlated domains can be much larger than
the size in equilibrium, growing with time as some power of time. This
conclusion has been supported by analogies with condensed matter
systems obeying classical  dissipative stochastic first order equations.
 However, as has beens stressed by Boyanovsky
 {\it et al.} [4] the situation in a relativistic field theory, where the
order parameter obeys a quantized
conservative second order equation does not necessarily have
to be the same. The same problem of determining the rate of growth of domain
structures out of equilibrium is also of importance in cosmology, since the
size of the domains determine the number of different topological defects
produced during the phase transition. In this paper, we report on an
attempt to study this phenomenon systematically within the context
of a model that exhibits dynamical symmetry breaking.

The method we will use to analyze this problem is the  Closed Time
Path formalism invented in the sixties by Schwinger and Keldysh [5]. Since
chiral
symmetry breaking is not well understood in QCD even at zero temperature, we
will take the Nambu-Jona-Lasinio [6] model as an effective theory having the
essential physics of chiral symmetry breaking present in QCD. In this model the
critical temperature is a function of the coupling constant $g$ and, instead of
coupling our system to some heat bath  to generate the quenching, we will
mimic the drop in temperature by suddenly changing $g$, and consequently the
critical temperature. Ideally one should study such a phenomenon
starting from an initial temperature $T$ (and coupling constant $g$ such that
$T>T_c(g)$ ) and then suddenly increasing $g$ to $g'$ at some instant
$t_0$ such that $T<T_c(g')$. The system, in this case, will find itself in
the wrong phase and the formation of domains in the broken phase will take
place. Our aim is to calculate the value of $<\bar\psi(t,\vec
r)\psi(t,\vec r) \bar\psi (t,0)\psi(t,0)>$ which gives a measure of the
distance scale over which isospin is correlated. However, such a
calculation is quite complicated and to bring out the qualitative features
we choose an alternate but parallel model where the interaction ($g$)
is turned on suddenly at $t=0$ such that the initial temperature
$T<T_c(g)$. We find that the domain size grows with time (for large times)
as $\sqrt{t}$ in a model independent manner.

{\bf II. Closed Time Path formalism}

In this section we briefly review the Closed Time Path formalism. All
essential details
can be found
in the references [5,7,8]. Suppose we are interested in the expectation
value of
some observable $A$ at instant $t$ in a given ensemble. It is given, in
general by
$$<A> (t) = Tr \rho (t) A = Tr \rho(0) U(0,t) A U(t,0),\eqno(1)$$
\noindent
where $\rho$ is the density matrix that describes the (mixed) state of the
system and $U$ is the evolution operator  (in the Schroedinger picture)
corresponding to a hamiltonian $H$ that is , in general, time dependent.
The density matrix does not necessarily have to commute with the hamiltonian,
in which case it describes a non equilibrium state. For $t\le 0$ we
assume that the system is in equilibrium so that at $t=0$, $\rho$ can be
written as
$$\rho(0)={e^{-\beta H_i}\over Tr   e^{-\beta H_i} }  ,\eqno(2)$$
\noindent
for some initial hamiltonian $H_i$.
For positive times we take $H$ as the hamiltonian which governs the
dynamics of the system. We can write now
$$\rho(0)={ e^{-\beta H_i}\over Tr e^{-\beta H_i} }=
{U(T-i\beta, T)\over Tr \ U(T-i\beta, T) },\eqno(3)$$
\noindent
where $T$ is any time $T<0$, since $U(T-i\beta, T)$ involves only the
hamiltonian $H_i$.
We can now go back to equation (1). After substituting equation (3),
 inserting the identity operators
$1=U(0, T) U(T,0)$, $1=U(t,T') U(T',t)$ and rearranging the product inside the
trace we obtain
$$<A>(t)={Tr U(T-i\beta, T) U(T,T') U(T',t) A U(t,T)\over
Tr U(T-i\beta, T) U(T,T') U(T',T)}   ,\eqno(4)$$
\noindent
where $T'$ is some time larger than $t$. Equation (4) can be pictured as
describing the evolution
of the system from $T$ to $T'$ (with an insertion of the operator $A$ at $t$),
and then backwards in time to $T$ and finally along the imaginary time
axis to $T-i\beta$.
We can now take  the limits $T\rightarrow -\infty $ and $T'\rightarrow
\infty$.  This suggests the definition of a generating functional
$$Z_c[J_c]=Tr\  U_J(-\infty-i\beta, -\infty) U_J(-\infty, \infty) U_J(\infty,
-\infty),\eqno(5)$$
\noindent
where the subcript $c$ is to remind us that the quantity is defined in the
contour in the complex $t$-plane described above.
$U_J$ stands for the evolution operator under the influence of the external
sources $J_c$. Clearly, if $H$ equals $H_i$ and is time independent, and
if the
source $J_c$ is the same along the entire time
contour, $Z_c$ is the partition
function of the
system and this formalism reduces to the well known imaginary time formalism.
We are, however, dealing with a nonequilibrium phenomenon where $H(t)$ is
different along the time contour and, consequently we choose the external
sources to be different along the two branches of
the contour to allow us to obtain Green's functions by taking derivatives
with respect to the sources.
  $Z_c$ has a path integral representation
$$Z_c[J_c]=\int_{\rm (anti)periodic}
D\phi \quad e^{i\int_c ({\cal L}+J\phi) d^4x}.\eqno(6)$$
\noindent
with $\phi$ denoting a generic field of the theory.
As usual we can separate
the
quadratic part of the lagrangian and expand the interaction part in a power
series to obtain a perturbative expansion. It turns out that the
free propagators
vanish if they connect a point in the segment ($-\infty, -\infty-i\beta$) to a
point on the real axis. Thus the contribution of this segment effectively
decouples and contributes only to an overall normalization [8]. Therefore, we
are effectively left
with an integral on the real axis (both ways).  It is
convenient to rewrite
this  as
a normal integral over the real line along one branch, but the propagator
then acquires a $2 \times 2$ matrix structure
$$\Delta_{ab}=\left \vert\matrix{\Delta_{++}   &\Delta_{+-}\cr
                               \Delta_{-+}  &\Delta_{--}\cr}
\right\vert\eqno(7) $$
\noindent
where
$$\eqalign{\Delta_{++}(x-y)&=<T(\phi(x) \phi(y))>_\beta\cr
           \Delta_{--}(x-y)&=<T^* (\phi(x)\phi(y))>_\beta\cr
           \Delta_{+-}(x-y)&=<\phi(y)\phi(x)>_\beta\cr
           \Delta_{-+}(x-y)&=<\phi(x)\phi(y)>_\beta .\cr}\eqno(8)$$
\noindent
Here $T$ and $T^*$ stand for time ordering and antitime ordering respectively.
The matrix structure and the different
boundary conditions satisfied by the functions $\Delta_{a b}$
can be understood
remembering that in the part of the contour that
runs in the negative direction (we call it "$-$" as opposed to the other
one that we call "$+$") the time ordering gets reversed, and that any point
in the "$+$" branch
is earlier than any point in the "$-$" branch. Besides the
matrix structure of the propagator the other difference in relation to the
usual Feynman rules at $T=0$ is that we have the interaction term in both
branches of the contour and
consequently there are two kinds of vertices, one connecting fields
living on the "$+$" branch and the other (with
opposite sign) connecting fields defined on the "$-$" branch.
\bigskip

\noindent
{\bf III. The Model}

The model we would like to consider is the Nambu-Jona-Lasinio model in four
dimensions, with two flavors and $N$ colors.
$${\cal L}=i\bar\psi_\alpha\gamma^\mu\partial_\mu \psi_\alpha
+ {g^2\over 4N
\Lambda^2} [(\bar\psi_\alpha\psi_\alpha)^2 - (\bar\psi_\alpha
\gamma_5\tau^i\psi_\alpha)^2], \eqno(9)$$
\noindent
where $\alpha=1, . . . , N$,
$\tau^i$'s are the generators of flavour $SU(2)$,
and the flavor indices are suppressed for
simplicity. This model is nonrenormalizable in four dimensions. However, we
treat this, for our discussions, as an effective low energy theory for energy
scales below the cut off $\Lambda$. The theory has the chiral symmetry
$$\eqalign{\psi_\alpha &\rightarrow e^{i\theta \gamma_5} \psi_\alpha\cr
  \bar{\psi}_\alpha
 &\rightarrow \bar\psi_\alpha e^{i\theta \gamma_5}.\cr}\eqno(10)$$
\noindent
As usual we can introduce auxiliary fields $\sigma$ and $\pi^i$ to
write the Lagrangian (9) in  an equivalent form
$${\cal L}=i\bar\psi_\alpha\gamma^\mu\partial_\mu  \psi_\alpha
-{N\Lambda^2\over 2}
\sigma^2 - {N\Lambda^2\over 2} \pi^i\pi^i
+ {g\over \sqrt{2}} \sigma \bar\psi_\alpha\psi_\alpha +
i {g\over \sqrt{2}} \pi^i \bar\psi_\alpha \gamma_5 \tau^i
\psi_\alpha.\eqno(11) $$
\noindent
As is well known, the chiral  symmetry (10) of the theory is dynamically
broken if $g^2>2 \pi^2$ and restored at
$T_c= {\rm const.} (1-{2\pi^2  \over g^2})^{1\over 2} \Lambda$.
In our discussion we will take the
coupling constant $g$ to be time dependent with the simple form
$$g(t)= \theta (t) g ,\eqno(12)$$
\noindent
and the initial temperature $T$ smaller than the critical temperature
$$T_c={\rm const.}(1-{2\pi^2\over g^2})^{1\over 2}\Lambda.\eqno(13)$$
\noindent
Thus, according to the model we have chosen,
for $t<0$ we have a gas of massless free
fermions in equilibrium at a temperature $T$ and the state of the system
is invariant under the chiral transformation of equation (10).
At $t=0$ we turn on
the interaction such
that the system suddenly finds itself at a temperature much below
 the critical temperature $T_c$ given in (13). This
guarantees that the effect of the sudden change will make the chiral
symmetric initial state of the system unstable, almost in parallel to the
case of an expanding plasma as discussed in the introduction.
 An important point
to note, however, is
that, since both the dynamics and the initial state are chirally symmetric,
$\sigma$ will have a vanishing vacuum expectation value in
the ensemble. However, in
each element of the ensemble,
different regions of the space will have different directions of
symmetry breaking and, although
the expectation value of $\sigma$ (averaged over the ensemble) remains zero,
correlations of the $\sigma$ (and $\pi$) field will grow.

The  complete $\sigma$
( or $\pi$ ) propagator is obtained from
$$D^{-1}_{a b}(x,y) = D^{0\ \ -1}_{a b} (x,y) + \Sigma_{a b}
(x,y),\eqno(14) $$
\noindent
where $a,b=\pm$, $D^0$ the zeroth order propagator and $\Sigma$ is   the
self-energy.
The  discussion of domain growth, however, is best carried out in
terms of the physical
propagators (namely, the retarded, advanced and correlated ones) and,
therefore, we transform from the $\pm$ basis to the physical basis
through the transformation [5,7],
$$D'=VDV^{-1},\ \ \ \ \Sigma'=V\Sigma V^{-1}    ,\eqno(15)$$ \noindent
with
$$V= {1\over \sqrt{2}}\left(\matrix{   1     &-1\cr
                                       1     &1\cr   }\right).\eqno(16)$$
\noindent
Thus,
$$D'=\left(\matrix{ 0    & -D_A\cr
                    D_R  &D_c\cr}\right),\eqno(17)$$
$$\Sigma'=\left(\matrix{\Sigma_c     &\Sigma_R\cr
                        -\Sigma_A  &0\cr}\right).\eqno(18)$$
\noindent
with
$$\eqalign{D_R &=D_{++}-D_{+-}\cr
           D_A &=-D_{++}+D_{-+}\cr
           D_c &=D_{++}+D_{--}\cr
           \Sigma_R &=\Sigma_{++}+\Sigma_{+-}\cr
           \Sigma_A &=-\Sigma_{++}-\Sigma_{-+}\cr
           \Sigma_c &=\Sigma_{++}+\Sigma_{--}.\cr}\eqno(19)$$
\noindent
The functions $D_{R(A)}$ are the usual retarded (advanced) functions
defined as
$$\eqalign{D_R(x,y)&=\theta(x^0-y^0) <[\phi(x),\phi(y)]>_\beta\cr
           D_A(x,y)&=\theta(y^0-x^0)<[\phi(x), \phi(y)]>_\beta ,\cr}\eqno(20)$$
\noindent
and $D_c$, which gives the fermionic correlation fuunction we are
interested in is
 $$D_c(x,y)= <[\phi(x), \phi(y)]_+>_\beta .\eqno(21)$$
\noindent
In case of Fermi  fields we would exchange commutators by anticommutators
and {\it vice versa}.
The difference in sign in  the definition of $\Sigma_{R(A)}$ and $D_{R(A)}$
is due to the fact that $\Sigma_{+-}$ and $\Sigma_{-+}$ include a
"$-$" sign coming from a "$-$" type vertex. Using equation (14) we find
the relation between the propagator and the self energy to be
$$D_R=D^0_R {1\over 1+\Sigma_R D^0_R},\eqno(22)$$
$$D_A=D^0_A{1\over 1+\Sigma_A D^0_A},\eqno(23) $$
\noindent
and
$$D_c=D_R \Sigma_c D_A + {1\over 1+D^0_R \Sigma_R}D_c^0
{1\over 1+\Sigma_AD^0_A} .\eqno(24)$$
\noindent
Note that in a non equilibrium situation these functions will depend on each
time argument separately, and not only on their difference.
In other words, in a non equilibrium situation there is no time translation
invariance. Thus, we can not
diagonalise them by taking a Fourier transform and the order of the functions
in (22, 23, 24) is important. Consequently, unlike the equilibrium case, it
is in
general difficult to invert expressions such as $1+\Sigma_R D^0_R$.
\bigskip

{\bf IV. The Calculation}
\noindent

The functions $D_{R(A)}$ carry information about the  dynamics of the model
and at zeroth order are independent of the initial state (they are
expectation values of c-numbers which vanish at equal times by
microcausality).  It is the function $D_c$ that contains
the information  about the state of the system, like particle number
distributions, temperature,etc. For instance, for free fermions we have
(We use the metric $\eta_{\mu\nu}=(+---)$)
$$\eqalign{S_{R(A)}(x)&=\int {d^4 k\over (2\pi)^4} e^{-ik.x}(k^\mu \gamma_\mu
+ m) {1\over k^2-m^2\pm i\epsilon k_0}\cr
           S_c(x)&=\int {d^4 k\over (2\pi)^4} e^{-ik.x}
(k^\mu\gamma_\mu+m) \quad2\pi i(2 n(k_0)-1) \delta(k^2-m^2),\cr}\eqno(25)$$
\noindent
where
$$n(k_0)={1\over e^{\beta |k_0|}+1}\eqno(26)$$
\noindent
is the Fermi distribution function.
We want to calculate $D_c$ for the $\sigma $ (and $\pi$) propagators at
equal times. At tree
level these are not real (on shell) propagating particles,
consequently they do not thermalise. Another way of saying this is that they
are too heavy (their masses are of the order of the cut-off) and they are
Boltzmann suppressed at any temperature much smaller than the cut-off.
We have then
$$\eqalign{D^0_R &={1\over iN\Lambda^2},\cr
           D^0_A &=-{1\over iN\Lambda^2},\cr
           D_c^0 &=0 .\cr}\eqno(27)$$
We will calculate the self-energy in the leading order in ${1\over N}$ so
that the main contribution comes from the fermionic one loop
bubble diagram. But we note first that since $g(t)=\theta(t)
g$, $\Sigma_{R(A)}(x,y)$ and $\Sigma_c(x,y)$ vanish if either
$x^0<0$ or $y^0<0$. Consequently, they have the generic form
$$\Sigma(x,y)=\theta(x^0) \bar\Sigma (x,y) \theta(y^0).\eqno(28)$$
\noindent
where the quantity with an overbar is calculated with a constant $g$ much
the same way as in an equilibrium calculation. As a result, the inversion
of quantities such as $(1+\Sigma_R D^0_R)^{-1}$ becomes trivial for
positive times. For example, $(1+\Sigma_R D^0_R)^{-1}$ satisfies the equation
 $$(1+\Sigma_R
D^0_R)^{-1}(1+\Sigma_R D^0_R)=1.\eqno(29)$$
\noindent
Written out explicitly, this has the form
$$\int d^4z\ (1+\Sigma_R D^0_R)^{-1} (x,z) [\delta^4(z-y) + {1\over
iN\Lambda^2}
\theta(z^0) \oSigma_R(z-y) \theta(y^0)]=\delta^4(x-y).\eqno(30)$$
\noindent
By definition $\bar\Sigma_R$ vanishes for $z^0<y^0$ and consequently, if
we restrict to $y^0>0$, the two step functions in the second term can be
ignored and the inversion can be done much the same way as for a constant
coupling, equilibrium case. The knowledge of $(1+\Sigma_R D^0_R)^{-1}$ for
positive times is enough to calculate the retarded propagator for positive
times ($x^0, y^0>0$)
$$\eqalign{D_R(x,y)&=\int dz\quad D^0_R(x,z) (1+\Sigma_R D^0_R)^{-1}(z,y)\cr
                   &=\oD_R (x-y).\cr}\eqno(31)$$
\noindent
The same argument goes through for the advanced function
as well but this is not
always
the case. For the Feynman Greens function, for example, the analogue of
equation (30) is a complicated integral equation. This comes about because
$\Sigma_{++}$ describes propagation in both, forward and backward time
direction
so, even after restricting the final points $x^0$ and $y^0$ to be positive,
$z^0$, the intermediate time coordinate,  can be negative. Since the dynamics
for negative times is differentfrom that
 at positive times the inversion is not straightforward. Since our
non equilibrium retarded and advanced
functions are the same as the ones in equilibrium  (always for
$x^0,y^0>0$ only) we have (It will become clear shortly that this is all
we need for our discussion.)
 $$\oD_R(x,y) = \int {d^4 k\over (2\pi)^4}
e^{-ik(x-y)}  {1\over iN\Lambda^2+\oSigma_R(k)},\eqno(32)$$
$$\oD_A(x,y)=\int {d^4 k\over (2\pi)^4} e^{-ik(x-y)}
{1\over -iN\Lambda^2+\oSigma_A(k)} .\eqno(33)$$
\noindent From equations (24) and (27) then, we obtain for $x^0,y^0>0$
$$\eqalign{D_c(x,y)&=\int d^4z d^4z' \quad D_R(x,z) \Sigma_c(z,z')
D_A(z',y)\cr
                 &=\int d^4z d^4z' \quad \oD_R(x-z) \theta(z^0)
\bar\Sigma_c(z-z')\theta({z'}^0) \oD_A(z'-y).\cr}\eqno(34)$$
\noindent
Note that to evaluate $D_c$ we need $D_{R(A)}$
only for positive times as stated earlier. To
find $\oSigma_R$,
$\oSigma_A$ and $\bar\Sigma_c$ we first calculate $\oSigma_{ab}$,
$a,b=\pm$ using the diagrammatic methods and then
use
the definitions in (19). $\oSigma_{ab}$ is simply obtained from  the Feynman
rules discussed above to be
$$\oSigma_{ab} = -ab \ \  g^2 N\  {\rm Tr}\int {d^4 k\over (2\pi)^4}
 \quad
iS_{ab}(k) iS_{ba}(p+k),\eqno(35)$$
\noindent
where $S_{ab}$ are the finite temperature, real time propagators at the
initial temperature $T$
$$\eqalign{ S_{++}(p)&=(\gamma^\mu p_\mu+m)\left( {1\over p^2-m^2+i\epsilon}+
2\pi i n(p_0)\delta(p^2-m^2)\right)\cr
            S_{--}(p)&=(\gamma^\mu p_\mu+m)\left(- {1\over p^2-m^2-i\epsilon}+
2\pi i n(p_0)\delta(p^2-m^2)\right )\cr
            S_{+-}(p)&=(\gamma^\mu p_\mu+m) 2\pi i
(n(p_0)-\theta(-p_0))\delta(p^2-m^2)\cr
            S_{-+}(p)&=(\gamma^\mu p_\mu+m) 2\pi i
(n(p_0)-\theta(p_0))\delta(p^2-m^2).\cr}\eqno(36)$$
\noindent
and "$ab$" is a sign factor coming from the fact that
the "-" vertices carry a negative sign.

Note that in our theory $m=0$ and using these definitions
the calculation of
$\oSigma_{ab}$ is straighforward. We only give the results here:
$$\eqalign{\oSigma_{++}&=i4g^2N\int\dkthree {1-2n(\omega_k)\over 2\omega_k}
\lbrace [
{p_0 \omega_k+\pk\over \pmenos}
-{i\pi\over\omega_k}(p_0\omega_k-\pk)\cr
&\phantom{N \dkthree 1-2 n \omega ))}
\delta(\pmais)({1\over 2}
sgn(p_0)+n(\omega_k)
-n(\omega_k)n(\omega_{p+k})) ]\cr
&\phantom{1666 \quad 4 g^2 1-n\omega \quad\quad}+ p_0\rightarrow -p_0\rbrace
\cr
\oSigma_{--}&=\oSigma_{++}^* (p^*)\cr}$$
$$\eqalign{\oSigma_{+-}&= 4 \pi g^2 N \int \dkthree {1\over \omega_k} [
(p_0\omega_k-\pk)\delta(\pmais)\cr
&\phantom{4 \pi g^2 N \int \dkthree {1\over 2}\quad}
(n(\omega_k)-1)(n(p_0+\omega_k)-\theta
(-p_0-\omega_k)) \cr
&\phantom{4\pi g^2 N \int \dkthree {1\over 2}\quad}
+(-p_0^2-\pk)\delta(\pmenos)\cr
&\phantom{4\pi g^2 N \int\dkthree {1\over 2}\quad}
n(\omega_k)(n(p_0-\omega_k)-\theta(-p_0+\omega))]\cr
 \oSigma_{-+}&=4\pi g^2 N\int \dkthree {1\over \omega_k} [
(p_0\omega_k-\pk)\delta(\pmais)\cr
&\phantom{4 \pi g^2 N\int \dkthree {1\over 2}\quad}
n(\omega_k)(n(p_0-\omega_k)
-\theta(-p_0+\omega_k))\cr
&\phantom{ 4 g^2 N \int \dkthree {1\over 2}}
+(-p_0^2-\pk)\delta(\pmenos)\cr
&\phantom{4 \pi g^2 N\int \dkthree {1\over 2}\quad}
(n(\omega_k)-1)(n(p_0+\omega_k)
-\theta(-p_0-\omega_k))
].\cr}\eqno(37)$$
\noindent
We can now construct $\oSigma_R$, $\oSigma_A$ and $\bar \Sigma_c$. They are
given,
after some rearrangement and with the help of the change of variables
$k\rightarrow -k-p$ in various places
 (these are finite integrals since the momenta are cut off at $\Lambda$), by
$$\eqalign{\oSigma_R(p)&=i4g^2N\int \dkthree ({1-2n(\omega_k)\over 2
\omega_k})
 [ {p_0\omega_k+\pk\over (p_0-\omega_k+i\epsilon)^2-\omega_{p+k}^2}+
p_0,\epsilon\rightarrow -p_0,-\epsilon ]\cr
\oSigma_A(p)&=\oSigma_R^*(p^*)\cr}\eqno(38)$$
\noindent
and
$$\eqalign{\bar\Sigma_c(p)=-4g^2\pi N (p_0^2-\vec p^2)\int \dkthree
{1\over\omega_k}[
&\delta(\pmais)({1\over 2}
sgn(p_0)+n(\omega_k)\cr
&-n(\omega_k)n(p_0+\omega_k))
+p_0\rightarrow p_0 ].\cr}\eqno(39)$$
\noindent
Consequently we can also obtain $\bar D_{R(A)}$ in a straightforward manner.
We will need to know the position of the pole(s)
$\epsilon_p$($\epsilon_p^*$) of
$\oD_{R(A)}(p_0,\vec p)$ in order to evaluate $D_c$. We determine the
position of the pole in the following way.
At the pole of $\bar D_R$ we have
  $$iN\Lambda^2+\oSigma_R(p_0, \vec p)=0.\eqno(40)$$
\noindent
If we set $\vec p=0$ in the above expression, we obtain
$$\Lambda^2\left(1-{g^2\over 2\pi^2}\right )+
{g^2\over 8\pi^2 }p_0^2 \ln ({-4\Lambda^2\over p_0^2})-{16 g^2\over \pi^2}
\int^\Lambda_0 dk
{1\over e^{\beta k}+1} {k^3\over p_0^2-4k^2}=0.\eqno(41)$$
\noindent
A simple graphical analysis shows that, for $g^2>2\pi^2$ (the symmetry
breaking case) and temperatures smaller than  $T_c$, there is always
a solution of equation (41) with $p_0^2$ negative. Let us call this solution
$-M^2$.
To find the next term in the series expansion we substitute $p_0^2\rightarrow
-M^2+b \vec p^2$ and keep terms up to second order in $\vec p$. After some
tedious algebra we obtain
$$({7\over 3} - b)I_3 + ({5\over 3} - b) I_5=0,\eqno(42)$$
\noindent
where
$$I_3=4M^4\int^\Lambda_0 dk\  {\rm tgh}(\beta k/2) {k^3\over
(M^2+4k^2)^3}\eqno(43)$$
\noindent
and
$$I_5=16M^2\int^\Lambda_0 dk\  {\rm tgh}(\beta k/2) {k^5\over
 (M^2+4k^2)^3}.\eqno(44)$$
\noindent
The solution to equation (42) can be easily seen to lie within the range
$${5\over 3}<b<{7\over 3}.\eqno(45) $$
\noindent
The position of the pole up to this order is given by $p_0^2=\epsilon_p^2=
-M^2+b \vec p^2$, with $b$ determined as above. This way we have
$$\epsilon_{\vec p}\sim \pm i(M-{b\vec p^2\over 2M}).\eqno(46)$$
\noindent
The wrong sign of the mass $(-M^2)$ is a signal
of instability of the initial state. One could go on to higher orders but this
will suffice for our discussion.

Using our previous results we have now, with $x=(t, \vec r)$ and
$y=(t,\vec 0)$
 $$\eqalign{D_c(x,y)&=\int d^4z d^4z' \theta(z^0)\theta(z'^0)\int {d^4 k\over
(2\pi)^4}
\dpfour \dq e^{-ik.(x-z)-ip.(z-z')-iq.(z'-y)}\cr
&\phantom {\int dkdkdkdkdkdkdkdkdk}\times \ \oD_R(k)\bar\Sigma_c(p)\oD_A(q)\cr
&=\int {d^4 k\over (2\pi)^4} {dp_0\over 2\pi}
{dq_0\over 2\pi}e^{i\kr}e^{i(k_0-q_0)t}
{i\over k_0-p_0-i\epsilon} {i\over p_0-q_0+i\epsilon}\cr
&\phantom{ dkdkdkdkdkdkdkdkdk}\times \ \oD_R(k_0,\vec k)
\bar\Sigma_c(p_0,\vec k) \oD_A(q_0, \vec k).\cr}\eqno(47)$$
\noindent
Note that it is the limited range of integrations on $z^0, z'^0$
that produces
the denominators shown above
instead of the usual $\delta$ functions that assure conservation of
energy.
As the system is unstable, $\oD_R$
grows exponentially with time so that one should be
careful in defining its Fourier transform. It is well known [9] that if $\oD_R
\sim e^{\alpha t}$ for large
$t$,  the Fourier transform $\oD_R(\omega)$ will be
 analytic only above the line
Im  $\omega=\alpha$ and
the inverse Fourier transform is given by an integral along a
line just above Im $\omega=\alpha$.
The opposite happens for the advanced
function, namely, the region of
analyticity is Im $ \omega < \alpha$. With this in mind we can perform the
integrations over $k_0$ and $q_0$ to obtain
$$\eqalign{\int {dk_0\over 2\pi} e^{-ik_0t}\oD_R(k_0,\vec k)
{1\over k_0-p_0+i\epsilon}=
&-i \od_R(\epsilon_k,\vec k) {e^{-i\epsilon_k t}\over \epsilon_k-p_0} -i
e^{-ip_0 t} \oD_R(p_0,\vec k)\cr
           \int {dq_0\over 2\pi} e^{iq^0 t}\oD_A(q_0,\vec k)
{1\over q_0-p_0-i\epsilon}=
&i \od_A(\epsilon_k^*,\vec k) {e^{i\epsilon^*_k t}\over \epsilon^* - p_0}+
i e^{ip_0 t} \oD_A(p_0,\vec k),\cr}\eqno(48)$$
\noindent
where $\od_{R(A)}$ are the residues of $\oD_{R(A)}$ at the pole
$\epsilon_k(\epsilon^*_k)$. As is clear from the earlier discussion,
$\epsilon_k$ is complex leading to both exponentially growing as well as
damped behaviour. For large times, it is the exponentially growing
solution which will dominate. Furthermore, the terms not involving
$\epsilon_k$ do
not grow with time and are the ones that would be there had $g$ been kept
constant and the range of
integration in $z^0, z'^0$ was ($-\infty, \infty$) (namely, for the case
of equilibrium). We will omit  this term from now on
since the exponential growth gives the dominant contribution.
Putting this back
in (47) we have
$$\eqalign{D_c(t,\vec r)&=\int {dp_0\over 2\pi} \dkthree e^{i \kr}
|\od_R(\epsilon_k, \vec k)|^2
e^{-i(\epsilon_k-\epsilon^*_k)t} {1\over |p_0-\epsilon_k|^2}
\bar\Sigma_c(p_0, \vec k)\cr
&= \int\dkthree e^{i\kr} e^{-i(\epsilon_k-\epsilon^*_k)t} f(\vec
k^2,T),\cr}\eqno(49)$$
\noindent
where
$$\eqalign{f(\vec{k}^2, T)=g^2 N\int& {d^3 p\over (2\pi)^3}
{|\bar d_R(\epsilon_k,\vec k)|^2\over
\omega_k\omega_{p+k}}[
{((\omega_k+\omega_{p+k})^2-\vec k^2)\over
|\omega_p+\omega_{p+k}+\epsilon_k|^2}
({1\over2} - n(\omega_p)+ n(\omega_p)n(\omega_{p+k}))\cr
&+{((\omega_k-\omega_{p+k})^2-\vec k^2)\over
|\omega_p-\omega_{p+k}-\epsilon_k|^2}
({1\over 2}
sgn(\omega_p-\omega_{p+k})-n(\omega_p)+
n(\omega_p) n(\omega_{p+k}))\cr
&+{((\omega_k-\omega_{p+k})^2-\vec k^2)\over
|\omega_{p+k}-\omega_p-\epsilon_k|^2}
({1\over2}
sgn(\omega_p-\omega_{p+k})-n(\omega_p)+ n(\omega_p)n(\omega_{p+k}))\cr
&+{((\omega_p+\omega_{p+k})^2-\vec k^2)\over
|\omega_p+\omega_{p+k}-\epsilon_k|^2}
(-{1\over 2} -
n(\omega_p)+n(\omega_p)n(\omega_{p+k}))].\cr}\eqno(50)$$
\noindent
A little analysis shows that the function $f$ vanishes for $\vec
k^2\rightarrow 0$ so, for small  $\vec k^2$ we can write ($k=|\vec k|$)
$$f(k^2, T)\rightarrow T^3 g(k/T),\eqno(51) $$
\noindent
with $g(x)\rightarrow 0$ as $x\rightarrow 0$ as a power, e.g. $g(x)\sim
x^\alpha,
\alpha >0$. We evaluate $D_c$ in equation (49)
 for large times using the
saddle point approximation. The saddle point is given by
$$k_s={i 2M r\over b t}.\eqno(52)$$
\noindent
Therefore, $k\rightarrow 0$ as $t\rightarrow\infty$ and we are
justified in disregarding
terms of order O($p^4$) in the expansion of $\epsilon_p$ in equation (46).
This gives for large times
$$\eqalign{D_c(t,r)&= C(T) \left ({r\over t}\right )^\alpha
e^{M t} e^{-{M r^2\over 2 b t}}\cr
&= C(T)\left ({r\over t}\right )^\alpha e^{M t} e^{-{r^2\over
L^2(t)}}\cr}\eqno(53)$$ \noindent
where $C(T)$ is a function independent of $(t,\vec r)$ and
$L(t)$ is the typical size of the domains given by
$$L(t)=\sqrt{ {b\over 2M}t}.\eqno(54)$$
\noindent
 This
is the same scaling as obtained by Boyanovsky {\it et al.} [4] for the scalar
$\phi^4$ theory with a mass square that suddenly changes sign.
The size of the domains, of course, depends on the details of the theory since
it is determined by $b$ and $M$ but the power law dependence on $t$ in
equation
(54) seems to depend only on the fact that the first non constant term in
the expansion of $\epsilon_p$ for small $\vec p$ is quadratic in $\vec p$.
We see from (53) that the correlation grows indefinetely with time but this is
an artifact of our approximation.
This growth will actually stop when inside a domain the value of $\sigma$
is that of the equilibrium and the phase transition will be
 over. This
however
would involve a rearrangement of the chiralities of the fermions that
can not be achieved in leading order in ${1\over N}$ since in this
case the fermions do not scatter.
\bigskip
\noindent
{\bf V. Conclusion}
\bigskip
We have shown how to use the Closed Time Path formalism to describe the
dynamics of a phase transition. In particular we analysed the dynamics
of domain growth
following   quenching in a simple model exhibiting  chiral symmetry
breaking. We found that the domains with correlated isospin grow
as $\sqrt{t}$  and we have argued that this result is
not dependent on details of
our model.

This work was supported in part by US DOE Grant No. DE--
FG02--91ER40685. We would
like to thank Prof.
A.Salam, the IAEA and UNESCO for hospitality at the International Centre
for Theoretical Physics where part of this work was done. One of us (P.B.)was
partially supported by CAPES.
\bigskip
\bigskip
{\bf References}
\bigskip
\noindent
[1] A.A.Anselm -{\it Phys.Lett.} B217 (1989) 169

\noindent
\phantom{[1]} A.A.Anselm and M.G.Ryskin - {\it Phys.Lett. } B266 (1991) 482

\noindent
\phantom{[1]} J.D.Bjorken -{\it Int.J.Mod.Phys.} A7 (1992) 4189

\noindent
\phantom{[1]} J.P.Blaizot and A.Krzywicki -{\it Phys.Rev.} D46 (1992 )246

\noindent
\phantom{[1]} K.L.Kowalsky and C.C.Taylor - Case Western Reserve University
Preprint 92-6

\noindent
\phantom{[1]} S.Yu.Khlebnikov -UCLA preprint UCLA/93/TEP/10, hep-ph/9305207

\noindent
\phantom{[1]} I.Kogan -Princeton University preprint PUPT-1401, hep-th/9305291

\noindent
[2] C.M.G.Lattes, Y.Fujimoto and S.Hasegawa - {\it Phys.Rep.} 65, 151 (1980)

\noindent
[3] K.Rajagopal anf F.Wilczek -Princeton preprint PUPT-1347, IASSNS-HEP-92/60

\noindent
[4] D.Boyanovsky, D.-S.Lee, A.Singh -Pittsburgh preprint PITT 92-07

\noindent
[5] J.Schwinger -{\it J.Math.Phys.} 2, 407 (1961)

\noindent
\phantom{[5]} L.V.Keldysh -{\it Sov.Phys.JETP} 20, 1018 (1965)

\noindent
[6] Y.Nambu and G.Jona-Lasinio -{\it Phys.Rev.} 122 (1961) 345

\noindent
[7] K.C.Chou, Z.B.Su, B.-C.Hao and L.Yu -{\it Phys.Rep.} 118, 1 (1985)

\noindent
[8] G.Semenoff and N.Weiss -{\it Phys.Rev} D31, 689 (1985)

\noindent
[9] J.Schwinger -Lecture Notes of Brandeis University Summer Institute in
Theoretical Physics (1960)
 \end